\documentclass[a4paper,10pt]{article}

\usepackage{epsfig}

\newcommand{\CTT}{C^{\rm TT}_\ell}
\newcommand{\CTE}{C^{\rm TE}_\ell}
\newcommand{\CEE}{C^{\rm EE}_\ell}
\newcommand{\CBB}{C^{\rm BB}_\ell}
\newcommand{\CTB}{C^{\rm TB}_\ell}
\newcommand{\CEB}{C^{\rm EB}_\ell}
\newcommand{\EI}{{\cal E}_{\rm Inf}}

\newcommand{\BOOMERANG}{Boom2K flight}

\newcommand{\mpc}{\ensuremath{{\rm\,Mpc}}}

\newcommand{\hmpc}{\ensuremath{h^{-1}{\rm\,Mpc}}}

\newcommand{\Omhh}{\Omega_mh^2}
\newcommand{\Obhh}{\Omega_bh^2}

\def\lsim{\mathrel{\rlap{\lower4pt\hbox{\hskip1pt$\sim$}}
    \raise1pt\hbox{$<$}}}                
\def\gsim{\mathrel{\rlap{\lower4pt\hbox{\hskip1pt$\sim$}}
    \raise1pt\hbox{$>$}}}                

\def\refcite{\cite}

\begin{document}

\markboth{Tarun Souradeep}
{Cosmological quests in the CMB sky}

%
%

\title{Cosmological quests in the CMB sky}

\author{Tarun Souradeep \\Inter-University Centre for Astronomy and
Astrophysics,\\ Post Bag 4, Ganeshkhind, Pune 411~007, India.\\
tarun@iucaa.ernet.in}

\date{} 
\maketitle 
\footnotetext[1]{Invited plenary talk at the
International Conference on {\it Einstein's Legacy in the New Millennium},
Dec.15-22,2005, Puri, India.}
\begin{abstract}
Observational Cosmology has indeed made very rapid progress in recent
years. The ability to quantify the universe has largely improved due
to observational constraints coming from structure formation
Measurements of CMB anisotropy and, more recently, polarization have
played a very important role. Besides precise determination of various
parameters of the `standard' cosmological model, observations have
also established some important basic tenets that underlie models of
cosmology and structure formation in the universe -- `acausally'
correlated initial perturbations in a flat, statistically isotropic
universe, adiabatic nature of primordial density perturbations.  These
are consistent with the expectation of the paradigm of inflation and
the generic prediction of the simplest realization of inflationary
scenario in the early universe. Further, gravitational instability is
the established mechanism for structure formation from these initial
perturbations. In the next decade, future experiments promise to
strengthen these deductions and uncover the remaining crucial
signature of inflation -- the primordial gravitational wave
background.
\end{abstract}


\section{Introduction}	
 
The `standard' model of cosmology must not only explain the dynamics
of the homogeneous background universe, but also satisfactorily
describe the perturbed universe -- the generation, evolution and
finally, the formation of large scale structures in the universe. It
is fair to say much of the recent progress in cosmology has come from
the interplay between refinement of the theories of structure
formation and the improvement of the observations.

The transition to precision cosmology has been spearheaded by
measurements of CMB anisotropy and, more recently,
polarization. Despite its remarkable success, the `standard' model of
cosmology remains largely tied to a number of fundamental assumptions
that have yet to find precise observational verification~: the
Cosmological Principle, the paradigm of inflation in the early
universe and its observable consequences (flat spatial geometry, scale
invariant spectrum of primordial seed perturbations, cosmic
gravitational radiation background etc.).  Our understanding of
cosmology and structure formation necessarily depends on the
relatively unexplored physics of the early universe that provides the
stage for scenarios of inflation (or related alternatives).  The CMB
anisotropy and polarization contains information about the
hypothesized nature of random primordial/initial metric perturbations
-- (Gaussian) statistics, (nearly scale invariant) power spectrum,
(largely) adiabatic vs.  iso-curvature and (largely) scalar vs. tensor
component.  The `default' settings in bracket are motivated by
inflation.  Estimation of cosmological parameters implicitly depend on
the assumed values of the initial conditions, or, explicitly on the
scenario of generation of initial perturbations~\cite{recons_us}.
Besides precise determination of various parameters of the `standard'
cosmological model, observations have also established some important
basic tenets of cosmology and structure formation in the universe --
`acausally' correlated initial perturbations, adiabatic nature
primordial density perturbations, gravitational instability as the
mechanism for structure formation. We have inferred a spatially flat
universe where structures form by the gravitational evolution of
nearly scale invariant, adiabatic perturbations in a predominant form
of non--baryonic cold dark matter which is sub-dominant to a form dark
energy that does not cluster (on astrophysical scales).

The signature of primordial perturbations on super-horizon scales at
decoupling in the CMB anisotropy and polarization are the most
definite evidence for new physics (eg., inflation ) in the early
universe that needs to be uncovered.  We briefly review the
observables from the CMB sky and importance to understanding cosmology
in section~\ref{cmb} The article only briefly touches on the important
aspect of precision cosmology and summarizes the recent estimates of
the cosmological parameters. This information is available in more
detail in many recent literature,
eg. Ref.\refcite{sper_wmap06,boom_polar}.  The main theme of the
article is to highlight \footnote{The article does {\em not} attempt
at a review and is far from being exhaustive in the coverage of the
science and literature.} the quest cosmological observations in
establishing some of the fundamental tenets of cosmology and
structure~:

\begin{itemize}

\item{} Statistical Isotropy of the universe (Sec.~\ref{SI});

\item{} Gravitational instability mechanism for structure
formation(Sec.~\ref{GI});

\item{} Primordial perturbations from Inflation.(Sec.~\ref{PI}).

\end{itemize}

At this time, the attention of the community is largely focused on
estimating the cosmological parameters.  The next decade would see
increasing efforts to observationally test fundamental tenets of the
cosmological model using the CMB anisotropy and polarization
measurements (and related LSS observations, galaxy survey,
gravitational lensing, etc.).

\section{CMB observations and cosmological parameters}
\label{cmb}

The angular power spectra of the Cosmic Microwave Background
temperature fluctuations ($C_\ell$)have become invaluable observables
for constraining cosmological models. The position and amplitude of
the peaks and dips of the $C_\ell$ are sensitive to important
cosmological parameters, such as, the relative density of matter,
$\Omega_0$; cosmological constant, $\Omega_\Lambda$; baryon content,
$\Omega_B$; Hubble constant, $H_0$ and deviation from flatness
(curvature), $\Omega_K$.

\begin{figure}
\begin{center}
\includegraphics[scale= 0.4,angle=-90]{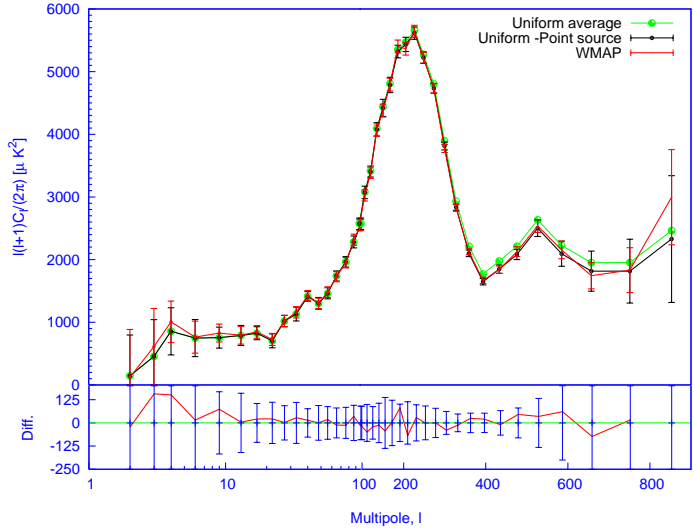}
\caption{\footnotesize The angular power spectrum estimated from the
WMAP multi-frequency using a self-contained model free approach to
foreground removal (black curve) is compared to the WMAP team estimate
(red).
 The published binned WMAP power spectrum plotted
in red line with error bars for comparison. The lower panel shows the
difference in the estimated power spectra. The method holds great
promise for CMB polarization where modeling uncertainties for
foregrounds are much higher.}
\label{WMAPCL}
\end{center}
\end{figure}

  The angular spectrum of CMB temperature fluctuations has been
measured with high precision on large angular scales ($\ell < 800$) by
the WMAP experiment \cite{hin_wmap06}, while smaller angular scales
have been probed by ground and balloon-based CMB experiments
\cite{ruh02,read04,dick04,kuo04,hal02}.  These data are broadly
consistent with a $\Lambda$CDM model in which the Universe is
spatially flat and is composed of radiation, baryons, neutrinos and,
the exotic, cold dark matter and dark energy. The exquisite
measurements by the Wilkinson Microwave Anisotropy Probe (WMAP) mark a
successful decade of exciting CMB anisotropy measurements and are
considered a milestone because they combine high angular resolution
with full sky coverage and extremely stable ambient condition (that
control systematics) allowed by a space mission .  Figure~\ref{WMAPCL}
shows the angular power spectrum of CMB temperature fluctuations
obtained from the first year of WMAP data~\cite{sah06}.  The third
year of WMAP observations have also included CMB polarization
results. The WMAP results are of excellent quality and show robustness
to different analysis methods~\cite{wmap3reanal}.

One of the firm predictions of this working `standard' cosmological
model is linear polarization pattern ($Q$ and $U$ Stokes parameters)
imprinted on the CMB at last scattering surface. Thomson scattering
generates CMB polarization anisotropy at decoupling~\cite{cmb_polar}.
This arises from the polarization dependence of the differential cross
section: $d\sigma/d\Omega\propto |\epsilon'\cdot\epsilon|^2$, where
$\epsilon$ and $\epsilon'$ are the incoming and outgoing polarization
states~\cite{rad_basics} involving linear polarization only.  A local
quadrupole temperature anisotropy produces a net polarization, because
of the $\cos^2\theta$ dependence of the cross section.  A net pattern
of linear polarization is retained due to local quadrupole intensity
anisotropy of the CMB radiation impinging on the electrons at
$z_{rec}$. The coordinate--free description decomposes the two kinds
of polarization pattern on the sky based on their different parities.
In the spinor approach, the even parity pattern is called the
$E$--mode and the odd parity pattern the $B$--mode.  With the
introduction of polarization, there are a total of 4 power spectra to
determine: $\CTT, \CTE, \CEE, \CBB$. Parity conservation ~\footnote{On
the other hand, a non-zero detection of $\CTB$ or $\CEB$, over and
above observational artifacts, could be tell-tale signatures of exotic
parity violating physics~\cite{parviol}.} eliminates the two other
possible power spectra, $\CTB$ \& $\CEB$. While CMB temperature
anisotropy can also be generated during the propagation of the
radiation from the last scattering surface, the CMB polarization
signal can be generated only at the last scattering surface, where the
optical depth transits from large to small values. The polarization
information complements the CMB temperature anisotropy by isolating
the effect at the last scattering surface from effects along the line
of sight.

\begin{figure}
\begin{center}
\includegraphics[scale=0.6]{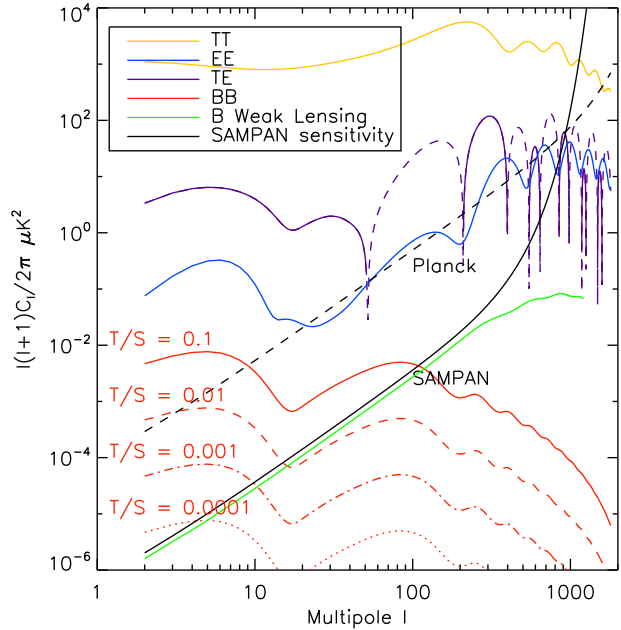}
\caption{Figure taken from Ref.~\protect{\refcite{bouch05}} shows the
CMB anisotropy (TT) and the three polarization power spectra
(TE,EE,BB) that can be extracted from CMB sky. The $\CBB$ spectra
being the weakest, presents the toughest challenge. The three $\CBB$
curves correspond to a specific energy scale of inflation ranging from
$2.2$ to $0.4$ in units of $(10^{16} GeV)$.  These predictions will be
within reach of dedicated satellite missions presently under study
both at ESA and NASA.}
\end{center}
\label{cmbpol}
\end{figure}

The CMB polarization is an even cleaner probe of early universe
scenarios, that promises to complement the remarkable successes of CMB
anisotropy measurements. As seen in figure~\ref{cmbpol}, the CMB
polarization signal is much smaller than the anisotropy
signal. Measurements of polarization at sensitivities of $\mu K$
(E-mode) to tens of $nK$ level (B-mode) pose spectacular challenges
for ongoing and future experiments.

 After the first detection of CMB polarization by DASI in 2003, the
field has rapidly grown, with measurements coming in from a host of
ground--based and balloon--borne dedicated CMB polarization
experiments. The Degree Angular Scale Interferometer (DASI) measured
the CMB polarization spectrum over a limited band of angular scales
($l\sim 200-440$) in late 2002~\cite{kov_dasi02}.  The DASI experiment
recently published results of much refined measurements with 3 years
of data~\cite{dasi_3y}. More recently, the Boomerang collaboration
reports measurements of $\CTT$, $\CTE$ and $\CEE$ and a non--detection
of $B$--modes~\cite{boom_polar}. The recent release of full sky E-mode
polarization maps and polarization spectra by WMAP are a new milestone
in CMB research~\cite{pag_wmap06,kog_wmap03}.  As expected, there has
been no detection of cosmological signal in B-mode of
polarization. The lack of $B$--mode power suggests that foreground
contamination is at a manageable level which is good news for future
measurements.  Scheduled for launch in 2007, the Planck satellite will
greatly advance our knowledge of CMB polarization by providing
foreground/cosmic variance--limited measurements of $\CTE$ and $\CEE$
out beyond $l\sim 1000$.  We also expect to detect the lensing signal,
although with relatively low precision, and could see gravity waves at
a level of $r\sim 0.1$. In the future, a dedicated CMB polarization
mission has been listed as a priority by both NASA (Beyond Einstein)
and ESA (Cosmic Vision) in the time frame 2015-2020.  These primarily
target the $B$-mode polarization signature of gravity waves, and
consequently, identify the viable sectors in the space of inflationary
parameters.
\begin{table}[h] 
\begin{center}
\caption{The table taken from ~Ref.\protect{\refcite{sper_wmap06}}
summarizes the estimated values of the cosmological parameters of the
$\Lambda$CDM Model. The best fit parameters correspond to the maximum
of the joint likelihoods of various combinations of CMB anisotropy and
large scale structure data.}  {\begin{tabular}{|c||c|c|c|c|} 
\hline &&&& \\
{Data combo.$\rightarrow$} &WMAP & WMAP& WMAP+ACBAR & WMAP + \\ &Only & +CBI+VSA &
+BOOMERanG &2dFGRS \\ {Parameters $\downarrow$} & & & & \\ \hline
\hline
 &&&& \\
100$\Omega_b h^2$ & 
\ensuremath{2.233^{+ 0.072}_{- 0.091} \mbox{ }} &
\ensuremath{2.212^{+ 0.066}_{- 0.084} \mbox{ }} &
\ensuremath{2.231^{+ 0.070}_{- 0.088} \mbox{ }} &
\ensuremath{2.223^{+ 0.066}_{- 0.083} \mbox{ }}  \\
$\Omega_m h^2 $ & 
\ensuremath{0.1268^{+ 0.0072}_{- 0.0095} \mbox{ }} &
\ensuremath{0.1233^{+ 0.0070}_{- 0.0086} \mbox{ }} &
\ensuremath{0.1259^{+ 0.0077}_{- 0.0095} \mbox{ }} &
\ensuremath{0.1262^{+ 0.0045}_{- 0.0062} \mbox{ }}  \\
$h$ & 
\ensuremath{0.734^{+ 0.028}_{- 0.038} \mbox{ }} &
\ensuremath{0.743^{+ 0.027}_{- 0.037} \mbox{ }} &
\ensuremath{0.739^{+ 0.028}_{- 0.038} \mbox{ }} &
\ensuremath{0.732^{+ 0.018}_{- 0.025} \mbox{ }}  \\
$A$ & 
\ensuremath{0.801^{+ 0.043}_{- 0.054} \mbox{ }} &
\ensuremath{0.796^{+ 0.042}_{- 0.052} \mbox{ }} &
\ensuremath{0.798^{+ 0.046}_{- 0.054} \mbox{ }} &
\ensuremath{0.799^{+ 0.042}_{- 0.051} \mbox{ }}  \\
$\tau$ & 
\ensuremath{0.088^{+ 0.028}_{- 0.034} \mbox{ }} &
\ensuremath{0.088^{+ 0.027}_{- 0.033} \mbox{ }} &
\ensuremath{0.088^{+ 0.030}_{- 0.033} \mbox{ }} &
\ensuremath{0.083^{+ 0.027}_{- 0.031} \mbox{ }}  \\
$n_s$ & 
\ensuremath{0.951^{+ 0.015}_{- 0.019} \mbox{ }} &
\ensuremath{0.947^{+ 0.014}_{- 0.017} \mbox{ }} &
\ensuremath{0.951^{+ 0.015}_{- 0.020} \mbox{ }} &
\ensuremath{0.948^{+ 0.014}_{- 0.018} \mbox{ }}  \\
&&&& \\\hline 
\hline &&&& \\
$\sigma_8$ & 
\ensuremath{0.744^{+ 0.050}_{- 0.060} \mbox{ }} &
\ensuremath{0.722^{+ 0.043}_{- 0.053} \mbox{ }} &
\ensuremath{0.739^{+ 0.047}_{- 0.059} \mbox{ }} &
\ensuremath{0.737^{+ 0.033}_{- 0.045} \mbox{ }} 
\\
$\Omega_m $ & 
\ensuremath{0.238^{+ 0.030}_{- 0.041} \mbox{ }} &
\ensuremath{0.226^{+ 0.026}_{- 0.036} \mbox{ }} &
\ensuremath{0.233^{+ 0.029}_{- 0.041} \mbox{ }} &
\ensuremath{0.236^{+ 0.016}_{- 0.024} \mbox{ }} \\
 &&&& \\
\hline
\end{tabular}\label{tab:lcdm_low}}
\end{center}
\end{table}

The measurements of the anisotropy in the cosmic microwave background
(CMB) over the past decade has led to `precision cosmology'.
Observations of the large scale structure in the distribution of
galaxies, high redshift supernova, and more recently, CMB
polarization, have provided the required complementary
information. The current up to date status of cosmological parameter
estimates from joint analysis of CMB anisotropy and Large scale
structure (LSS) data is usually best to look up in the parameter
estimation paper accompanying the most recent results announcement of
a major experiment, such as recent WMAP release~\cite{sper_wmap06}.
Using WMAP data only, the best fit values for cosmological parameters
for the power-law, flat, $\Lambda$CDM model are $(\Omega_m h^2,
\Omega_b h^2, h, n_s, \tau, \sigma_8) =$ $(0.127^{+0.007}_{-0.013},
0.0223^{+0.0007}_{-0.0009}$, $0.73^{+0.03}_{-0.03}$,
$0.951^{+0.015}_{-0.019}$, $0.09^{+0.03}_{-0.03}$,
$0.74_{-0.06}^{+0.05})$.  Table~\ref{tab:lcdm_low} summarizes best fit
parameters that correspond to the maximum of the joint likelihoods (in
a multi-dimensional parameter space) of various combinations of CMB
anisotropy and large scale structure data.

\section{Statistical Isotropy of the universe}
\label{SI}

The {\em Cosmological Principle} that led to the idealized FRW universe
found its strongest support in the discovery of the (nearly)
isotropic, Planckian, Cosmic Microwave Background. The isotropy around
every observer leads to spatially homogeneous cosmological models.
The large scale structure in the distribution of matter in the
universe (LSS) implies that the symmetries incorporated in FRW
cosmological models are to be interpreted statistically.

The CMB anisotropy and its polarization is currently the most
promising observational probe of the global spatial structure of the
universe on length scales near to and even somewhat beyond the
`horizon' scale ($\sim c H_0^{-1}$).  The exquisite measurement of the
temperature fluctuations in the CMB provide an excellent test bed for
establishing the statistical isotropy (SI) and homogeneity of the
universe. In `standard' cosmology, CMB anisotropy signal is expected to
be statistically isotropic, i.e., statistical expectation values of
the temperature fluctuations $\Delta T(\hat q)$ are preserved under
rotations of the sky. In particular, the angular correlation function
$C(\hat{q},\, \hat{q}^\prime)\equiv\langle\Delta T(\hat q)\Delta
T(\hat q^\prime)\rangle$ is rotationally invariant for Gaussian
fields. In spherical harmonic space, where $\Delta T(\hat q)=
\sum_{lm}a_{lm} Y_{lm}(\hat q)$, the condition of {\em statistical
isotropy} (SI) translates to a diagonal $\langle a_{lm} a^*_{l^\prime
m^\prime}\rangle=C_{l} \delta_{ll^\prime}\delta_{mm^\prime}$ where
$C_l$, is the widely used angular power spectrum of CMB anisotropy.
The $C_l$ is a complete description of (Gaussian) SI CMB sky CMB
anisotropy and hence would be (in principle) an inadequate measure for
comparing models when SI is violated~\cite{bps}.

Interestingly enough, the statistical isotropy of CMB has come under a
lot of scrutiny after the WMAP results. Tantalizing evidence of SI
breakdown (albeit, in very different guises) has mounted in the {\it
WMAP} first year sky maps, using a variety of different statistics. It
was pointed out that the suppression of power in the quadrupole and
octopole are aligned \cite{maxwmap}.  Further ``multipole-vector''
directions associated with these multipoles (and some other low
multipoles as well) appear to be anomalously correlated
\cite{cop04,schw04}.  There are indications of asymmetry in the power
spectrum at low multipoles in opposite hemispheres
\cite{erik04a}. Possibly related, are the results of tests of
Gaussianity that show asymmetry in the amplitude of the measured genus
amplitude (at about $2$ to $3\sigma$ significance) between the North
and South galactic hemispheres \cite{erik04b,erik04c,par04}. Analysis
of the distribution of extrema in {\it WMAP} sky maps has indicated
non-gaussianity, and to some extent, violation of SI
\cite{lar_wan04}. The three-year WMAP maps are consistent with the
first-year maps up to a small quadrupole difference. The two
additional years of data and the improvements in analysis has not
significantly altered the low multipole structures in the
maps~\cite{hin_wmap06}. Hence, `anomalies' are expected to persist at
the same modest level of significance and are unlikely to be artifacts
of noise, systematics, or the analysis in the first year data.  The
cosmic significance of these `anomalies' remains debatable also
because of the aposteriori statistics employed to ferret them out of
the data. More importantly, what is missing is a common, well defined,
mathematical language to quantify SI (as distinct from non
Gaussianity) and the ability to ascribe statistical significance to
the anomalies unambiguously.

\begin{figure}[h]
\begin{center}
\includegraphics[scale=0.4, angle=-90]{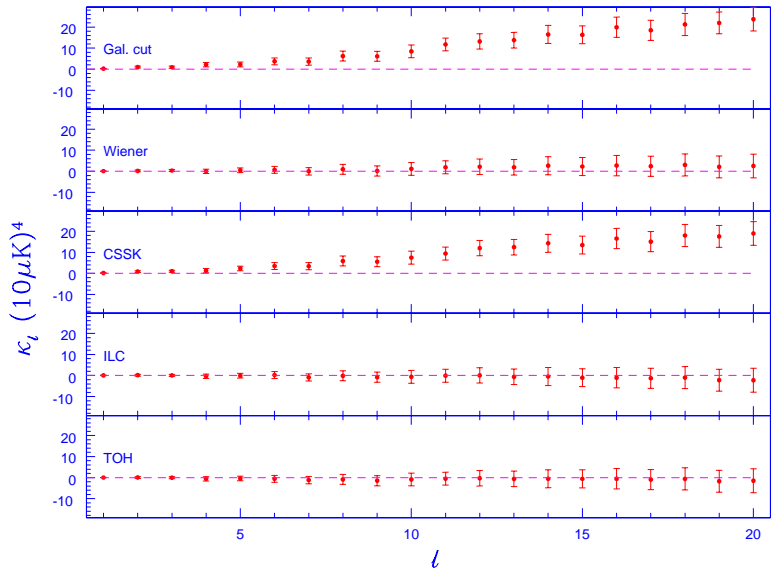}
 \caption{BiPS measurements for different CMB maps based on the
WMAP-1year data filtered with a window with $l_s =30$, $l_t =20$. This
measures the statistical isotropy of the WMAP in the modest $20<l<40$
range in the multipole space where certain anomalies have been
reported.  The bottom two panels show that BiPS of WMAP-1 is
consistent with SI. ILC with a $10^\circ$ galactic cut (top) has the
same BiPS as CSSK ($l_s =30$, $l_t =20$) and explains that the raising
tail of CSSK map is because of the mask. Although the Weiner filtered
map shows a null BiPS in this filter, it shows a rising tail just like
CSSK in a filter peaking a slightly larger $l$. This reflects the
measurable SI violation created by Weiner filtering that the
suppresses power in the foreground contaminated regions along a band
around the galactic plane~\protect{\cite{us_apj}}.}
\end{center}
\label{kappa_wmap_3020}
\end{figure}

The observed CMB sky is a single realization of the underlying
correlation, hence the detection of SI violation, or correlation
patterns, pose a great observational challenge. It is essential to
develop a well defined, mathematical language to quantify SI (as
distinct from non Gaussianity) and the ability to ascribe statistical
significance to the anomalies unambiguously.
Recently, the Bipolar Power spectrum (BiPS) $\kappa_\ell$
($\ell=1,2,3, \ldots$) of the CMB map was proposed as a statistical
tool of detecting and measuring departure from
SI~\cite{us_apjl,us_pascos}. The non-zero value of the BiPS spectrum
imply the break down of statistical isotropy
\begin{equation} 
{\mathrm { STATISTICAL\,\,\,\, ISOTROPY}} \,\,\,\,\,\,\, \Longrightarrow \,\,\,\,\,\,\, 
\kappa_\ell\,=\,0 \,\,\,\,\,\,\, \forall \ell \ne 0.
\end{equation}
BiPS is sensitive to structures and patterns in the underlying total
two-point correlation function \cite{us_apjl,us_pascos}.  The BiPS is
particularly sensitive to real space correlation patterns (preferred
directions, etc.) on characteristic angular scales. In harmonic space,
the BiPS at multipole $\ell$ sums power in off-diagonal elements of
the covariance matrix, $\langle a_{lm} a_{l'm'}\rangle$, in the same
way that the `angular momentum' addition of states $l m$, $l' m'$ have
non-zero overlap with a state with angular momentum
$|l-l'|<\ell<l+l'$. Signatures, like $a_{lm}$ and $a_{(l+n)\mbox{ }
m}$ being correlated over a significant range $l$ are ideal targets
for BiPS. These are typical of SI violation due to cosmic topology and
the predicted BiPS in these models have a strong spectral signature in
the bipolar multipole $\ell$ space~\cite{us_prl}.  The orientation
independence of BiPS is an advantage since one can obtain constraints
on cosmic topology that do not depend on the unknown specific
orientation of the pattern ({\it{e.g.}}, preferred directions).
Measurement of the BiPS on the following CMB anisotropy maps based the
first year WMAP data~: A) a foreground cleaned map (denoted as
`TOH')~\cite{maxwmap}; B) the Internal Linear Combination map (denoted
as `ILC' in the figures)~\cite{ben_wmap03}, and C) a customized linear
combination of the QVW maps of WMAP with a galactic cut (denoted as
`CSSK'). Fig.~\ref{kappa_wmap_3020} shows that the measured BiPS for
all the WMAP sky maps are consistent with statistical
isotropy~\cite{us_apjl2,us_apj}. The ongoing BIPS analysis on WMAP-3yr
data indicates that BiPS of the three years maps show an improvement
in SI -- the deviations are smaller and fewer~\cite{haj_sour06}.

CMB polarization maps over large areas of the sky have been recently
delivered by experiments in the near future. The statistical isotropy
of the CMB polarization maps will be an independent probe of the
cosmological principle.  Since CMB polarization is generated on at the
surface of last scattering, violations of statistical isotropy are
pristine cosmic signatures and more difficult to attribute to the local
universe.  The Bipolar Power spectrum has been defined and implemented
for CMB polarization and show great promise~\cite{bas06}.


BiPS is a promising probe to detect the topology of the universe.The
underlying correlation patterns in the CMB anisotropy in a multiply
connected universe is related to the symmetry of the Dirichlet
domain. The BiPS expected in flat, toroidal models of the universe has
been computed and shown to be related to the principle directions in
the Dirichlet domain \cite{us_prl}. As a tool for constraining cosmic
topology, the BiPS has the advantage of being independent of the
overall orientation of the Dirichlet domain with respect to the
sky. Hence, the null result of BiPS have important implication for
cosmic topology. This approach complements other direct search for
signature of cosmic topology~\cite{staro,circles} and our results are
consistent with the absence of the matched circles and the null S-map
test of the WMAP CMB maps~\cite{circles04,angelwmap,circles06}. Full Bayesian
likelihood comparison to the data of specific cosmic topology models
is another approach that has applied to COBE-DMR data~\cite{bps}. The
BiPS has also been used to constrain anisotropic models of cosmology
using SI violating CMB patterns that arise in them~\cite{ghos06}.

\section{Gravitational instability mechanism for structure formation}
\label{GI}

It is a well accepted notion that the large scale structure in the
distribution of matter in the present universe arose due to
gravitational instability from the same primordial perturbation seen
in the CMB anisotropy at the epoch of recombination. This fundamental
assumption in our understanding of structure formation has recently
found an irrefutable direct observational
evidence~\cite{eis_sdss05,col_2Df05}.

The acoustic peaks occur because the cosmological perturbations excite
acoustic waves in the relativistic plasma of the early universe
\cite{peeb_yu70,sun_zel70,bon_efs84,bon_efs87,hol89}.  The
recombination of baryons at redshift $z\approx 1100$ effectively
decouples the baryon and photons in the plasma abruptly switching off
the wave propagation.  In the time between the excitation of the
perturbations and the epoch of recombination, modes of different
wavelength can complete different numbers of oscillation periods.
This translates the characteristic time into a characteristic length
scale and produces a harmonic series of maxima and minima in the CMB
anisotropy power spectrum.  The acoustic oscillations have a
characteristic scale known as the sound horizon, which is the comoving
distance that a sound wave could have traveled up to the epoch of
recombination.  This physical scale is determined by the expansion
history of the early universe and the baryon density that determines
the speed of acoustic waves in the baryon-photon plasma.

\begin{figure}[h]
\begin{center}
\includegraphics[scale=0.4]{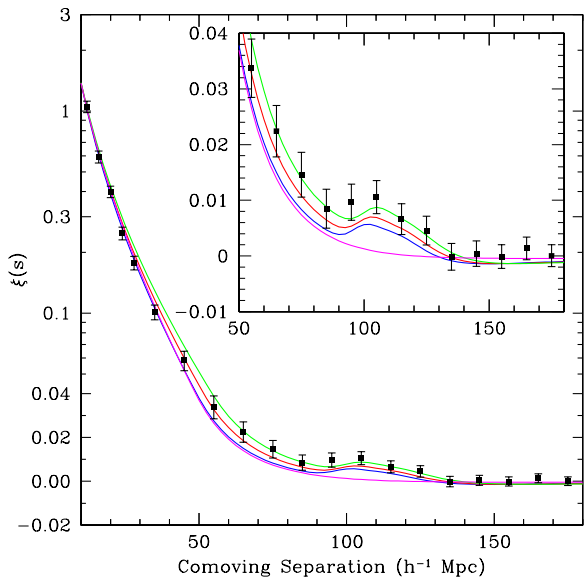}
\caption{ The large-scale redshift-space correlation function of the
SDSS LRG sample taken from Ref.~\protect{\refcite{eis_sdss05}}.  
The inset shows an expanded view with a linear vertical axis.  The
magenta line shows a pure CDM model ($\Omhh=0.105$), which lacks the
acoustic peak.  The models are $\Omhh=0.12$ (top, green), 0.13 (red),
and 0.14 (bottom with peak, blue), all with $\Obhh=0.024$ and $n=0.98$
and with a mild non-linear prescription folded in.  The clearly
visible bump at $\sim 100\hmpc$ scale is statistically significant.}
\end{center}
\label{fig:xi} 
\end{figure}

For baryonic density comparable to that expected from Big Bang
nucleosynthesis, acoustic oscillations in the baryon-photon plasma
will also be observably imprinted onto the late-time power spectrum of
the non-relativistic matter.  This is easier understood in a real
space description of the response of the CDM and baryon-photon fluid
to metric perturbations~\cite{eis_sdss05}. An initial small
delta-function (sharp spike) adiabatic perturbation ($\delta \ln
a|_H$) at a point leads to corresponding spikes in the distribution of
cold dark matter (CDM), neutrinos, baryons and radiation (in the
`adiabatic' proportion, $1+w_i$, of the species).  The CDM
perturbation grows in place while the baryonic perturbation being
strongly coupled to radiation is carried outward in an expanding
spherical wave.  At recombination, this shell is roughly $105
h^{-1}\mpc$ in (comoving) radius when the propagation of baryons
ceases.  Afterward, the combined dark matter and baryon perturbation
seeds the formation of large-scale structure.  The remnants of the
acoustic feature in the matter correlations are weak ($10\%$ contrast
in the power spectrum) and on large scales. The acoustic oscillations
of characteristic wavenumber translates to a bump (a spike softened by
gravitational clustering of baryon into the well developed dark matter
over-densities) in the correlation function at $105 h^{-1}\mpc$
separation.  The large-scale correlation function of a large
spectroscopic sample of luminous, red galaxies (LRGs) from the Sloan
Digital Sky Survey that covers $\sim 4000$ square degrees out to a
redshift of $z\sim 0.5$ with $\sim 50,000$ galaxies has allowed a
clean detection of the acoustic bump in distribution of matter in the
present universe.  Figure~\ref{fig:xi} shows the correlation function
derived from SDSS data that clearly shows the acoustic `bump' feature
at a fairly good statistical significance~\cite{eis_sdss05}. The
acoustic signatures in the large-scale clustering of galaxies provide
direct, irrefutable evidence for the theory of gravitational
clustering, notably the idea that large-scale fluctuations grow by
linear perturbation theory from $z\sim 1000$ to the present due to
gravitational instability.

\section{Primordial perturbations from Inflation}
\label{PI}

Any observational comparison based on the structure formation in the
universe necessarily depends on the assumed initial conditions
describing the primordial seed perturbations.  It is well appreciated
that in `classical' big bang model the initial perturbations would
have had to be generated `acausally'. Besides resolving a number of
other problems of classical Big Bang, inflation provides a mechanism
for generating these apparently `acausally' correlated primordial
perturbations~\cite{inflpert}.

 The power in the CMB temperature anisotropy at low multipoles
($l\lsim 60$) first measured by the COBE-DMR~\cite{cobedmr} did
indicate the existence of correlated cosmological perturbations on
super Hubble-radius scales at the epoch of last scattering, except for
the (rather unlikely) possibility of all the power arising from the
integrated Sachs-Wolfe effect along the line of sight. Since the
polarization anisotropy is generated only at the last scattering
surface, the negative trough in the $C_l^{TE}$ spectrum at $l\sim 130$
(that corresponds to a scale larger than the horizon at the epoch of
last scattering) measured by WMAP first sealed this loophole, and
provides an unambiguous proof of apparently `acausal' correlations in
the cosmological perturbations~\cite{pag_wmap06,kog_wmap03,ben_wmap03}. 

Besides, the entirely theoretical motivation of the paradigm of
inflation, the assumption of Gaussian, random adiabatic scalar
perturbations with a nearly scale invariant power spectrum is arguably
also the simplest possible choice for the initial perturbations.  What
has been truly remarkable is the extent to which recent cosmological
observations have been consistent with and, in certain cases, even
vindicated the simplest set of assumptions for the initial conditions
for the (perturbed) universe discussed below.

\subsection{Nearly zero curvature of space}

The most interesting and robust constraint obtained in our quests in
the CMB sky is that on the spatial curvature of the universe. The
combination of CMB anisotropy, LSS and other observations can pin down
the universe to be flat, $\Omega_K \approx-0.02\pm 0.02$. This is
based on the basic geometrical fact that angular scale subtended in
the sky by the acoustic horizon would be different in a universe with
uniform positive (spherical), negative (hyperbolic), or, zero
(Euclidean) spatial curvature.  Inflation dilutes the curvature of the
universe to negliglible values and generically predicts a (nearly)
Euclidean spatial section.

The CMB data~\cite{boom_polar} alone places a constraint on the
curvature which is $\Omega_k = -0.037^{+0.033}_{-0.039}$.  Addition of
the LSS data, yields a median value of $\Omega_k = -0.027 \pm 0.016$.
Restricting $H_0$ by the application of a Gaussian HST prior, the
curvature density determined from the \BOOMERANG\ data set and all
previous CMB results was $\Omega_k = -0.015 \pm 0.016$.  A constraint
$\Omega_k =-0.010\pm 0.009$ obtained by combining CMB data with the
red luminous galaxy clustering data, which has its own signature of
baryon acoustic oscillations \cite{eis_sdss05}. The WMAP 3 year data
can (jointly) constrain $\Omega_k =-0.024^{+0.016}_{-0.013}$ even when
allowing for dark energy with arbitrary (constant) equation state
$w$~\cite{sper_wmap06}. (The corresponding joint limit from WMAP-3yr
on the equation of state is also impressive,
$w=-1.062^{+0.128}_{-0.079}$).

\subsection{Adiabatic primordial perturbation}

The polarization measurements provides an important test on the
adiabatic nature primordial scaler fluctuations~\footnote{ Another
independent observable is the baryon oscillation in LSS discussed in
sec~\ref{GI}}.  CMB polarization is sourced by the anisotropy of the
CMB at recombination, $z_{rec}$, the angular power spectra of
temperature and polarization are closely linked.  Peaks in the
polarization spectra are sourced by the velocity term in the same
acoustic oscillations of the baryon-photon fluid at last
scattering. Hence, a clear indication of the adiabatic initial
conditions is the compression and rarefaction peaks in the temperature
anisotropy spectrum be `out of phase' with the gradient (velocity)
driven peaks in the polarization spectra.

\begin{figure}[h]
\begin{center}
  \includegraphics[scale=0.4, angle=-90]{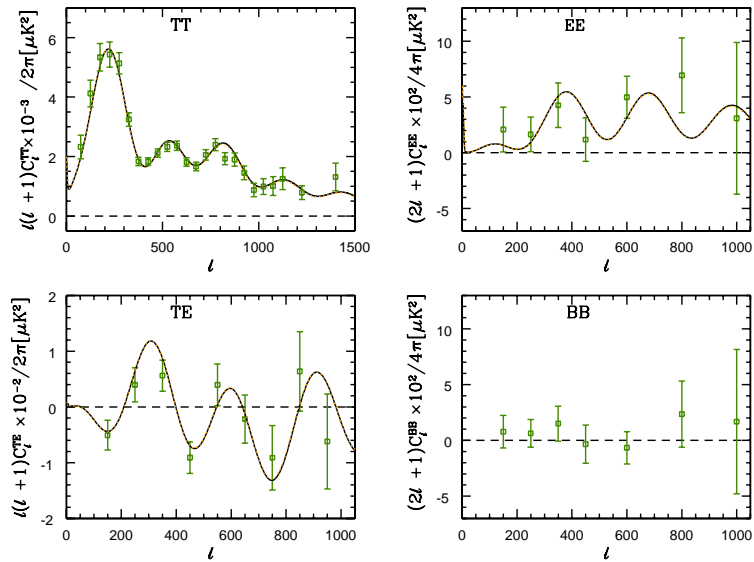}
\caption{Figure taken from Ref.~\protect{\refcite{boom_polar}} show
the measurement of the angular power spectrum CMB anisotropy and
polarization from Boomerang balloon 2000 flight. The CMB anisotropy
(TT) spectra has been measured at larger as well smaller angular
scales.  Data from other experiments such as DASI, CAPMAP and CBI are
comparable. The data is good enough to indicate that the peaks in EE
and TE are out of phase with that of TT as expected for adiabatic
initial conditions. The null BB detection of primary CMB signal from
gravity waves is not unexpected (given the ratio of tensor to scalar
perturbations) but contains good tiding for level of foregrounds and
the ability to deal with it.}
\label{boom_cmbspec}
\end{center}
\end{figure}

The figure~\ref{boom_cmbspec} taken from Ref.~\refcite{boom_polar} on
the Boomerang 2000 flight data reflects the current observational
status of CMB E-mode polarization measurements. The recent
measurements of the angular power spectrum the E-mode of CMB
polarization at large $l$ have confirmed that the peaks in the spectra
are out of phase with that of the temperature anisotropy
spectrum. Data from other experiments such as DASI, CAPMAP and CBI are
comparable. The data is good enough to indicate that the peaks in EE
and TE are out of phase with that of TT as expected for adiabatic
initial conditions~\cite{boom_polar}. The null BB detection of primary
CMB signal from gravity waves is expected given the ratio of tensor to
scalar perturbations but contain good tidings for the level of
foregrounds and the ability to deal with it.  These conclusions are
further borne out in the recent polarization results from the three
years of WMAP data~\cite{pag_wmap06}.

\subsection{Nearly scale-invariant power spectrum ?}
 
In a simple power law parametrization of the primordial spectrum of
density perturbation ($|\delta_k|^2 = A k^{n_s}$), the scale invariant
spectrum corresponds to $n_s=1$. Recent estimation of (smooth)
deviations from scale invariance favor a nearly scale invariant
spectrum~\cite{sel04}.  Current observations favor a value of $n_s =
0.98\pm0.02 \,(99.9\%CL)$ very close to unity are consistent with a
nearly scale invariant power spectrum.

The current combined CMB and LSS data is good enough to constrain the
`running' of the spectral index, $\alpha= d n_s/d\ln k = 0.003\pm
0.01\,(99.9\%CL)$. These results are remarkably consistent with the
generic predictions of the simplest models of inflation.  While the
simplest inflationary models predict that the spectral index varies
slowly with scale, inflationary models can produce strong scale
dependent fluctuations.  The first year WMAP observations provided
some motivation for considering these models as the data, particularly
when combined with the (first version of) measurements of the power
spectrum at much larger wavenumbers from the distribution of
absorption lines (Lyman-$\alpha$ forest) in the spectra of distant
quasars, were better fit by models with running spectral index.
Subsequent reanalysis of revised Lyman-$\alpha$ based measurements
showed the spectral index to be close to unity with no hint of
running~\cite{sel04}.


\begin{figure}
\begin{center}
\includegraphics[scale=0.4]{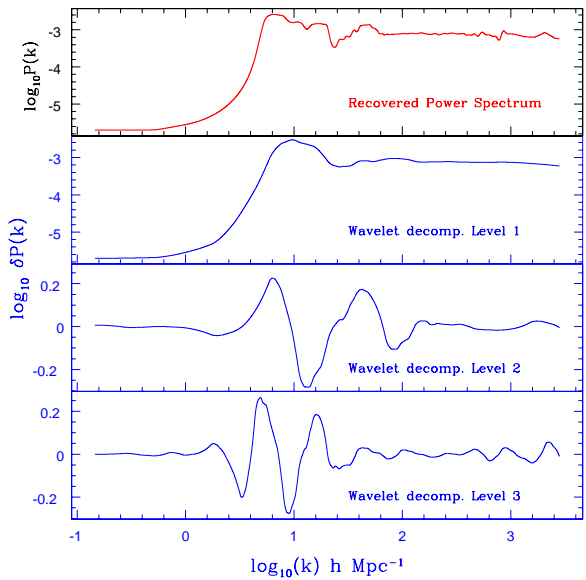}
\caption{The primordial power spectrum recovered from the angular
power spectrum of CMB anisotropy measured by WMAP is shown in the top
panel~\protect{\cite{shaf_sour04}}.  
Wavelet decomposition allows for clean
separation of the `features' in the recovered power spectrum on
different scales.  The lower three panels show a Daubechies-4 wavelet
decomposition from ongoing research work (Manimaran, Panigrahi,
Rangarajan, Souradeep).  The most significant feature at the coarsest
resolution is a compensated infrared cutoff shown in the second panel
(from top). The statistical significance of the small superimposed
oscillations is under closer study.}
\label{recovspec}
\end{center}
\end{figure}

Many model-independent searches have also been made to look for
features in the CMB power
spectrum~\cite{bridle03,hanne04,pia03,pia05}.  Accurate measurements
of the angular power spectrum over a wide range of multipoles from the
WMAP has opened up the possibility to deconvolve the primordial power
spectrum for a given set of cosmological
parameters~\cite{max_zal02,mat_sas0203,shaf_sour04,bump05}.  The
primordial power spectrum has been deconvolved from the angular power
spectrum of CMB anisotropy measured by WMAP using an improved
implementation of the Richardson-Lucy algorithm~\cite{shaf_sour04}.
The most prominent feature of the recovered primordial power spectrum
shown in Figure~\ref{recovspec} is a sharp, infra-red cut off on the
horizon scale. It also has a localized excess just above the cut-off
which leads to great improvement of likelihood over the simple
monotonic forms of model infra-red cut-off spectra considered in the
post WMAP literature.  The form of infra-red cut-off is robust to
small changes in cosmological parameters.  Remarkably similar form of
infra-red cutoff is known to arise in very reasonable extensions and
refinement of the predictions from simple inflationary scenarios, such
as the modification to the power spectrum from a pre-inflationary
radiation dominated epoch or from a sharp change in slope of the
inflaton potential~\cite{sin_sour04}.

\subsection{Gaussian primordial perturbations}
\label{gauss}

The detection of primordial non-Gaussian fluctuations in the CMB would
have a profound impact on our understanding of the physics of the
early universe. The Gaussianity of the CMB anisotropy on large angular
scales directly implies Gaussian primordial
perturbations~\cite{mun95,sper_gol99} that is theoretically motivated by
inflation~\cite{inflpert}. The simplest inflationary models predict
only very mild non-Gaussianity that should be undetectable in the
WMAP data.

The CMB anisotropy maps (including the non Gaussianity analysis
carried out by the WMAP team on the first year data~\cite{kom_wmap03})
have been found to be consistent with a Gaussian random field.
Consistent with the predictions of simple inflationary theories, no
significant deviations from Gaussianity in the CMB maps using general
tests such as Minkowski functionals, the bispectrum, trispectrum in
the three year WMAP data~\cite{sper_wmap06}. There have however been
numerous claims of anomalies in specific forms of non-Gaussian signals
in the CMB data from WMAP at large scales (see discussion in
sec.~\ref{SI}).

\subsection{Primordial tensor (GW) perturbations}

Inflationary models can produce tensor perturbations from
gravitational waves that are predicted to evolve independently of the
scalar perturbations, with an uncorrelated power spectrum.  The
amplitude of a tensor mode falls off rapidly on sub-Hubble radius
scales. The tensor modes on the scales of Hubble-radius the line of
sight to the last scattering distort the photon propagation and
generate an additional anisotropy pattern predominantly on the largest
scales. It is common to parameterize the tensor component by the ratio
$r_{k_*} = A_{\rm t}/A_{\rm s}$, ratio of $A_{\rm t}$, the primordial
power in the transverse traceless part of the metric tensor
perturbations, and $A_{\rm s}$, the amplitude scalar perturbation at a
comoving wavenumber, $k_*$ (in $\mpc^{-1}$).  For power-law models,
recent WMAP data alone puts an improved upper limit on the tensor to
scalar ratio, \ensuremath{r_{0.002} < 0.55 \mbox{ } (95\%\mbox{\ CL})}
and the combination of WMAP and the lensing-normalized SDSS galaxy
survey implies \ensuremath{r_{0.002} < 0.28 \mbox{ } (95\%\mbox{\
CL})} ~\cite{boom_polar}.

\begin{figure}[h]
\begin{center}
\includegraphics[scale=0.4]{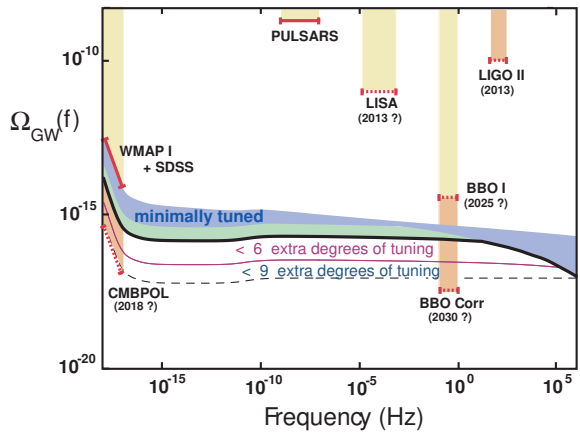} 
\caption{The figure taken from Ref.~\protect{\refcite{boy_stein06}}
shows the theoretical predictions and observational constraints on
primordial gravitational waves from inflation. The gravitational wave
energy density per logarithmic frequency interval, (in units of the
critical density) is plotted versus frequency. The blue region
represents the range predicted for simple inflation models with the
minimal number of parameters and tunings. The dashed curves have lower
values of tensor contribution, $r$, that is possible with more fine
tuned inflationary scenarios.  The currently existing experimental
constraints shown are due to: big bang nucleosynthesis (BBN), binary
pulsars, and WMAP-1 (first year) with SDSS. Also shown are the
projections for LIGO (both LIGO-I, after one year running, and
LIGO-II); LISA; and BBO (both initial sensitivity, BBO-I, and after
cross-correlating receivers, BBO-Corr). Also seen the projected
sensitivity of a future space mission for CMB polarization (CMBPol).}
\label{SGWBspec} 
\end{center}
\end{figure}

On large angular scales, the curl component of CMB polarization is a
unique signature of tensor perturbations.  The CMB polarization is a
direct probe of the energy scale of early universe physics that
generate the primordial metric perturbations.  Inflation generates
both (scalar) density perturbations and (tensor) gravity wave
perturbations. The relative amplitude of tensor to scalar
perturbations, $r$, sets the energy scale for inflation $\EI =
3.4\times 10^{16}$~GeV~$r^{1/4}$.  A measurement of $B$--mode
polarization on large scales would give us this amplitude, and hence
{\em a direct determination of the energy scale of inflation.}
Besides being a generic prediction of inflation, the cosmological
gravity wave background from inflation would be a fundamental test of
GR on cosmic scales and the semi--classical behavior of gravity.
Figure~\ref{SGWBspec} summarizes the current theoretical
understanding, observational constraints and future possibilities for
the stochastic gravity wave background from Inflation.

\section{Conclusions}

The past few years has seen the emergence of a `concordant'
cosmological model that is consistent both with observational
constraints from the background evolution of the universe as well that
from the formation of large sale structures.  It is certainly fair to
say that the present edifice of the `standard' cosmological models is
robust. A set of foundation and pillars of cosmology have emerged and
are each supported by a number of distinct observations~\cite{me_jpo}.

The community is now looking beyond the estimation of parameters of a
working `standard' model of cosmology. There is increasing effort
towards establishing the basic principles and assumptions.  The
feasibility and promise of this ambitious goal is based on the grand
success in the recent years with the CMB anisotropy measurements. The
quest in the CMB sky from ground, balloon and space have indeed
yielded great results!  While the ongoing WMAP and up coming Planck
space missions will further improve the CMB polarization measurements,
there are already proposals for the next generation dedicated
satellite mission in 2015-20 for CMB polarization measurements at best
achievable sensitivity.

\section*{Acknowledgments}

I would like to thank the organizers for arranging an excellent
scientific meeting at a lovely location. It is a pleasure to thank and
acknowledge the  contributions of students and collaborators
in the cosmological quest in the CMB sky at IUCAA.

\end{document}